\documentclass[12pt,draftclsnofoot, onecolumn]{IEEEtran}

\usepackage{amssymb}
\usepackage[cmex10]{amsmath}
\usepackage{stfloats}
\usepackage{graphicx}
\usepackage{subfigure}
\usepackage{tabularx}
\usepackage{epsfig,epsf,color,balance,cite}
\usepackage{verbatim}
\usepackage{url}
\usepackage{bm}

\newtheorem{coro}{\bf Corollary}
\newtheorem{theorem}{\bf Theorem}

\newtheorem{lemma}{\bf Lemma}

\usepackage{algorithm}
\usepackage{algorithmic}

\usepackage{color}
\definecolor{myc1}{rgb}{0,0,0}

\begin{document}

\title{\LARGE{Optimization of Rate Allocation and Power Control for Rate Splitting Multiple Access~(RSMA)}}

\author{
\IEEEauthorblockN{Zhaohui Yang, \IEEEmembership{Member, IEEE},
                  Mingzhe Chen, \IEEEmembership{Student Member, IEEE},
                  Walid Saad, \IEEEmembership{Fellow, IEEE},
                  and Mohammad Shikh-Bahaei, \IEEEmembership{Senior Member, IEEE}
                  \vspace{-3em}
                  }
\thanks{Part of the material in this paper was submitted to IEEE Globecom 2019 \cite{rsma2019}.}
\thanks{Z. Yang  and M. Shikh-Bahaei are with the Centre for Telecommunications Research, Department of Informatics, King’s College London, WC2B 4BG, UK, Emails: yang.zhaohui@kcl.ac.uk, m.sbahaei@kcl.ac.uk.}
\thanks{M. Chen is with Beijing Key Laboratory of Network System Architecture and Convergence,
Beijing University of Posts and Telecommunications, Beijing, China 100876, Email: chenmingzhe@bupt.edu.cn.}
\thanks{W. Saad is with Wireless@VT, Bradley Department of Electrical and Computer Engineering, Virginia Tech, Blacksburg, VA, USA, Email: walids@vt.edu.}
 }


\maketitle

\begin{abstract}
In this paper, the sum-rate maximization problem is studied for wireless networks that use downlink rate splitting multiple access (RSMA). In the considered model, each base station (BS) divides the messages that must be transmitted to its users into a ``private'' part and a ``common'' part. Here, the common message is a message that all users want to receive and the private message is a message that is dedicated to only a specific user. The RSMA mechanism enables a BS to adjust the split of common and private messages so as to control the interference by decoding and treating interference as noise and, thus optimizing the data rate of users. To maximize the users' sum-rate, the network can determine the rate allocation for the common message to meet the rate demand, and adjust the transmit power for the private message to reduce the interference. This problem is formulated as an optimization problem whose goal is to maximize the sum-rate of all users. To solve this nonconvex maximization problem, the optimal power used for transmitting the  private message to the users is first obtained in closed form for a given rate allocation and common message power. Based on the optimal private message transmission power, the optimal rate allocation is then derived under a fixed common message transmission power. Subsequently, a one-dimensional search algorithm is proposed to obtain the optimal solution of common message transmission power. Simulation results show that the RSMA can achieve up to 15.6\% and 21.5\% gains in terms of data rate compared to  non-orthogonal multiple access (NOMA) and orthogonal frequency-division multiple access (OFDMA), respectively.
\end{abstract}

\IEEEpeerreviewmaketitle

\section{Introduction}
Driven by the rapid development of  advanced multimedia applications such as virtual reality \cite{8395443}, next-generation wireless networks \cite{saad2019vision} must support high spectral efficiency and massive connectivity. {By splitting users in the power domain, non-orthogonal multiple access (NOMA) can simultaneously serve multiple users at the same
frequency or time resource \cite{liu2018non,vaezi2018multiple,7263349,Zhiguo2017Survey,Yang2018Power,8485386}. Consequently, NOMA-based
access scheme can achieve higher spectral efficiency than conventional orthogonal multiple access \cite{8422457,saito2013non,Yang2017On}. }
 However, using NOMA, the users must decode all of the interference as they receive the messages \cite{saito2013non}, which significantly increases the computational complexity needed for signal processing. To solve this problem, the idea of rate splitting multiple access (RSMA) was proposed in \cite{485709,1056307,7470942}. In RSMA, the message transmitted to the users is divided into a common message and a private message. The common message is a message that all users must receive and the private message is a message that only a specific intended user wishes to receive. To receive the common message, the users must decode the interference from other users. In contrast, to receive the private message, {\color{myc1}{the users must only consider the interference from other users' private messages which can be treated as noise.}} Therefore, adjusting the split of common and private messages can control the computational complexity and the data rate achieved by RSMA. However, implementing RSMA in wireless networks also face several challenges  \cite{7470942} such as the split of common and private message, resource management for effective private message transmission, and synchronization of message transmission.

Recently, a number of existing works such as in \cite{7470942,4039650,mao2018rate,mao2019rate,7555358,rahmati2019energy,7513415,7152864} studied important problems related to RSMA. The work in \cite{7470942} introduced the challenges and opportunities of using RSMA for multiple input multiple output (MIMO) based wireless networks. In \cite{4039650}, the authors proposed a distributed rate splitting method to maximize the data rates of the users. The authors in \cite{mao2018rate} evaluated the performance of RSMA, NOMA, and space-division multiple access (SDMA) and showed that RSMA achieves better performance than NOMA and SDMA. The authors in \cite{mao2019rate}  investigated the use of linearly-precoded rate-splitting method for simultaneous wireless information and power transfer networks. In \cite{7555358}, the authors used RSMA to maximize the rate of all users in downlink multi-user multiple input single output (MISO) systems under imperfect channel state information at the transmitter. The work in \cite{rahmati2019energy} studied the energy efficiency of the RSMA and NOMA schemes in a millimeter wave downlink transmission scenario. The use of RSMA is investigated in \cite{7513415} for a downlink multiuser MISO system with bounded errors in the channel state information at the transmitter. The authors in \cite{7152864} analyzed the data rate of using RSMA for two-receiver MISO broadcast channel with finite rate feedback. However, most of the existing works such as in  \cite{7470942,4039650,mao2018rate,mao2019rate,7555358,rahmati2019energy,7513415,7152864} that only find the suboptimal power control solutions for 
RSMA in different wireless systems such as MIMO and MISO, do not find optimal power control and rate allocation solutions for RSMA in single-input single-output (SISO) systems.
Meanwhile, none of these existing works \cite{7470942,4039650,mao2018rate,mao2019rate,7555358,rahmati2019energy,7513415,7152864} considers a successive interference cancelation (SIC) constraint for the private message transmission in RSMA,  which is needed to guarantee the successful decoding of the common message.


The main contribution of this paper is an optimized rate allocation and power control scheme for RSMA in  a downlink SISO system. To our best knowledge, this is the first work that finds, jointly, the optimal rate allocation solution for common message transmission and the optimal power allocation for both common and private message transmission. Our key contributions include:
\begin{itemize}
\item We propose a wireless network that uses RSMA and in which one base station (BS) transmits message to multiple users using RSMA scheme. To maximize the data rate of the users, we optimize the allocation of the data rate of common message transmission for each user as well as the transmit power that is used to transmit the common and private messages.
\item We formulate the considered rate allocation and power control problem as an optimization problem whose goal is to maximize the network sum-rate under both rate and SIC constraints.
To solve this problem, we first derive a closed-form expression for the optimal transmit power of the private message. Our fundamental analysis shows that, with the exception of one of the users, all  users are allocated with the minimum power to maintain the minimum rate demand.
Then, we characterize the finite solution space for the optimal rate allocation.
In order to obtain the optimal rate allocation and power control, {\color{myc1}{we propose a one-dimensional search algorithm that is shown to have linear complexity for equal rate demand.}}
\item 
Simulation results show that the optimized RSMA algorithm can achieve up to 15.6\% and 21.5\% gains in terms of data rate compared to NOMA and orthogonal frequency-division multiple access (OFDMA).
\end{itemize}
The rest of this paper is organized as follows. The system model and problem formulation are described in Section \uppercase\expandafter{\romannumeral2}. The optimal solution is presented in Section \uppercase\expandafter{\romannumeral3}. Simulation results are analyzed in Section \uppercase\expandafter{\romannumeral4}. Conclusions are drawn in Section \uppercase\expandafter{\romannumeral5}.

\section{System Model and Problem Formulation}

Consider the downlink a single-cell wireless network that consists of  one BS servicing a set $\mathcal K$ of $K$ users using RSMA \cite{485709}.
In RSMA, the common message is decoded by all users, while the individual private message is only decoded by each user.
At the receiver side, each user first decodes the common message and then decodes its private message using the previously decoded common message.

Let the common message of all users be $s_0$ and the private message of each user $k$ be $s_k$.
The transmitted signal $x$ of the BS is expressed as:
\begin{equation}\label{sys1eq1}
x=\sqrt{p_0}s_0  + \sum_{k=1}^K \sqrt{p_k}s_k,
\end{equation}
where $p_0$ is the transmit power of the common message $s_0$ and $p_k$ is the transmit power of the private message $s_k$ transmitted to user $k$.

The total received message at user $k$ can be given by:
\begin{equation}\label{sys1eq2}
y_k =\sqrt{h_k} x +n_k=\sqrt{h_kp_0}s_0  +\sum_{j=1}^K \sqrt{h_kp_j} s_j+n_k,
\end{equation}
where $h_k$ represents the channel gain between user $k$ and the BS and
 $n_k$ is the additive white Gaussian noise with variance $\sigma^2$.
The achievable rate of user $k$ decoding common message $s_0$ can be expressed as:
\begin{equation}\label{sys1eq3}
c_k =
B \log_2 \left( 1+ \frac{h_kp_0}{h_k\sum_{j=1}^Kp_j+\sigma^2}
\right),
\end{equation}
where $B$ is the bandwidth of the BS.
Without loss of generality, the channel gains are sorted in ascending order, i.e., $h_1\leq h_2\leq\cdots\leq h_K$.
To ensure that all users can successfully decode common message $s_0$, the rate of common message should be chosen as \cite{mao2018rate}:
\begin{align}\label{sys1eq5}
\min_{k\in\mathcal K}c_k
&=\min_{k\in\mathcal K}B \log_2 \left( 1+ \frac{p_0}{\sum_{j=1}^Kp_j+\frac{\sigma^2}{h_k}}
\right)\nonumber\\
&=B \log_2 \left( 1+ \frac{p_0}{\sum_{j=1}^Kp_j+\frac{\sigma^2}{\min_{k\in\mathcal K}h_k}}
\right)\nonumber\\
&\overset{\text{(a)}}{=}B \log_2 \left( 1+ \frac{p_0}{\sum_{j=1}^Kp_j+\frac{\sigma^2}{h_1}}
\right)\nonumber\\
&=c_1,
\end{align}
where equality (a) follows from the fact that $h_1\leq h_2\leq\cdots\leq h_K$.

To successfully implement SIC
 operation at the receiver, the transmit power of each user must satisfy the following constraint \cite{ali2016dynamic}:
\begin{equation}\label{sys1eq7_2}
{h_k}p_0 -{h_k}\sum_{j=1}^K p_j- { \sigma^2} \geq \theta, \quad \forall k\in\mathcal K,
\end{equation}
where $\theta$ is the minimum difference between the decoding signal power and the non-decoded inter-user interference  signal power plus noise power  \cite{8352643}.
This minimum difference is required to distinguish the common message to be
decoded and the remaining non-decoded private message of all users (plus noise).
Based on the channel condition $h_1\leq h_2\leq\cdots\leq h_K$, constraint \eqref{sys1eq7_2} can be simplified as:
\begin{equation}\label{sys1eq7_3}
p_0 -\sum_{j=1}^K p_j \geq  \frac{\theta+\sigma^2}{h_1}.
\end{equation}

Given the common message rate $c_1$ and the rate $a_k$ allocated to user $k$, the constraint of each user $k$'s data rate of receiving common message is given by:
\begin{equation}\label{sys1eq6}
\sum_{k=1}^Ka_k\leq c_1,
\end{equation}
where \eqref{sys1eq6} indicates that the total data rates of all users receiving common message must be less than the rate of common message $c_1$.

After having decoded the common message $s_0$, each user can decode its private message, the achievable rate of user $k$  decoding its private message $s_k$ is given by:
\begin{align}\label{sys1eq7}
r_k &=
B \log_2 \left( 1+ \frac{h_kp_k}{h_k\sum_{j=1,j\neq k}^Kp_j+\sigma^2}
\right).
\end{align}

Given the common message rate $a_k$ and achievable private message rate $r_k$, the total transmission rate of user $k$ in RSMA is:
\begin{equation}\label{sys1eq8}
r_{k}^{\text{tot}} =
a_k+r_k
=a_k+B \log_2 \left( 1+ \frac{h_kp_k}{h_k\sum_{j=1,j\neq k}^Kp_j+\sigma^2}
\right).
\end{equation}


%

\subsection{Problem Formulation}
Given the considered system model,  our objective is to optimize the rate allocation and power control so as to maximize the sum-rate under a total power constraint
and individual minimum rate requirements. Mathematically,
the sum-rate maximization problem for RSMA  can be formulated as:
\begin{subequations}\label{sys1min1}
\begin{align}
\mathop{\max}_{\boldsymbol a, \boldsymbol  p }\quad&\sum_{k=1}^K{\left(a_k+B \log_2 \left( 1+ \frac{h_kp_k}{h_k\sum_{j=1,j\neq k}^Kp_j+\sigma^2}\right)\right)}, \tag{\ref{sys1min1}}\\
\textrm{s.t.}\quad \:
&\sum_{k=1}^Ka_k\leq B \log_2 \left( 1+ \frac{h_1p_0}{h_1\sum_{j=1}^Kp_j+\sigma^2}
\right),\\
&{a_k+B \log_2 \left( 1+ \frac{h_kp_k}{h_k\sum_{j=1,j\neq k}^Kp_j+\sigma^2}\right)}\geq R_k,\quad\forall k \in \mathcal K,\\
&p_0 -\sum_{j=1}^K p_j \geq \frac{ \theta+\sigma^2}{h_1},\\
&\sum_{k=0}^Kp_k\leq P,\\
&a_k, p_0, p_k\geq 0,\quad \forall k \in \mathcal K,
\end{align}
\end{subequations}
where $\boldsymbol a=[a_1,a_2,\cdots,a_K]$, $\boldsymbol p=[p_0, p_1, p_2,\cdots, p_K]$,
$R_k$ is the minimum rate  demand of user $k$,
 and  $P$ is the maximum transmit power of the BS.
Constraint (\ref{sys1min1}a) ensures that each user can decode the common message.
The minimum rate constraints for all users are given in (\ref{sys1min1}b).
Constraint (\ref{sys1min1}c) shows the successful SIC power requirement, and (\ref{sys1min1}d) presents the maximum power constraint.

{\color{myc1}{
Since the objective function is not concave,
the sum-rate maximization problem (\ref{sys1min1}) is nonconvex.
Moreover, the rate and power vectors are coupled in the objective function and constraints, and hence, it is generally hard to solve problem (\ref{sys1min1}).
Despite the nonconvexity and coupling of variables in problem (\ref{sys1min1}), the globally optimal solution to problem (\ref{sys1min1}) can be effectively obtained in Section III.

 RSMA can potentially improve the rate of the network as shown in \cite{485709}. It is therefore imperative to quantify the performance gains that RSMA can obtain compared to conventional multiple access schemes. However, only suboptimal resource allocation is obtained for the sum-rate optimization with RSMA in the existing literature \cite{7470942,4039650,mao2018rate,mao2019rate,7555358,rahmati2019energy,7513415,7152864}. Although the sum-rate maximization problem in (\ref{sys1min1}) investigates SISO, the optimal solution can be derived, as shown in the following section, and this solution can serve as a benchmark for the optimization with multiple antenna.
}}

\section{Optimal Rate Allocation and Power Control}

\begin{figure}[t]
\centering
\includegraphics[width=5in]{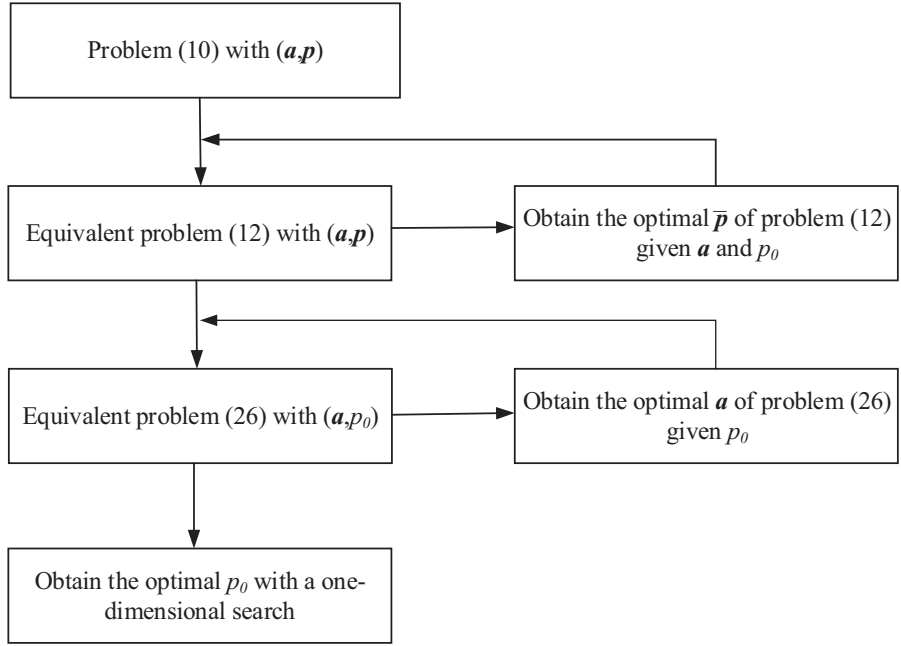}
\vspace{-1em}
\caption{Proposed approach for solving problem (\ref{sys1min1}).} \label{strcuter}
\vspace{-1em}
\end{figure}

In this section, we first provide the optimal conditions of problem (\ref{sys1min1}).
Then, based on these optimal conditions, the optimal private message transmission power is obtained in closed form under a given rate allocation and common message transmission power.
Substituting the optimal private message transmission power in problem (\ref{sys1min1}), the optimal closed-form rate allocation is then derived under a fixed common message transmission power.
Finally, a one-dimensional search algorithm is proposed to obtain the optimal solution of problem (\ref{sys1min1}).
The proposed process for solving problem (\ref{sys1min1}) is summarized in Fig. \ref{strcuter}.

\subsection{Optimal Conditions}

Before solving problem (\ref{sys1min1}), we provide some optimal conditions, which will be used to simplify problem (\ref{sys1min1}).
\begin{lemma}
At the optimal solution $(\boldsymbol a^*, \boldsymbol p^*)$ of problem (\ref{sys1min1}), the common message constraint (\ref{sys1min1}b) holds with equality, i.e., $\sum_{k=1}^Ka_k^*=\log_2 \left( 1+ \frac{h_1p_0^*}{h_1\sum_{j=1}^Kp_j^*+\sigma^2}
\right)$.
\end{lemma}

Lemma 1 can be easily proved by contradiction.

\begin{lemma}
At the optimal solution $(\boldsymbol a^*, \boldsymbol p^*)$ of problem (\ref{sys1min1}), the maximum power constraint (\ref{sys1min1}d) holds with equality, i.e., $\sum_{k=0}^Kp_k^*=P$.
\end{lemma}

\itshape {Proof:}  \upshape
See Appendix A.
\hfill $\Box$

Applying Lemma 1 and
substituting $\sum_{j=1,j\neq k}^Kp_j=P-p_0-p_k$ from Lemma 2 to (\ref{sys1min1}), we can then observe that problem (\ref{sys1min1}) is equivalent to the following problem:
\begin{subequations}\label{al2min1}
\begin{align}
\mathop{\max}_{\boldsymbol a, \boldsymbol  p }\quad&\sum_{k=1}^Ka_k
+\sum_{k=1}^K B\log_2 \left(\frac{h_k(P-p_0)+\sigma^2}{h_k(P-p_0-p_k)+\sigma^2}\right),
\tag{\theequation}\\
\textrm{s.t.}\quad \:
&\sum_{k=1}^Ka_k =    B\log_2 \left(\frac{h_1P  +\sigma^2}{h_1(P-p_0)+\sigma^2}\right),\\
&a_k+ B\log_2 \left(\frac{h_k(P-p_0)+\sigma^2}{h_k(P-p_0-p_k)+\sigma^2}\right)
  \geq R_k,
 \forall k \in \mathcal K,\\
&\sum_{k=0}^Kp_k = P,\\
&p_0\geq \frac P 2  + \frac{\theta +\sigma^2}{2h_1},\\
&a_k, p_k\geq 0, \quad \forall k \in \mathcal K.
\end{align}
\end{subequations}

To solve problem \eqref{al2min1}, we can show that it is further equivalent to another optimization problem, which admits a closed-form solution for the optimal private message transmission power.
\begin{lemma}
The optimal solution of problem \eqref{al2min1} is equivalent to the following problem:
\begin{subequations}\label{al2min1_2}
\begin{align}
\mathop{\max}_{\boldsymbol a,  {\boldsymbol p} }\quad& B\log_2 \left(\frac{h_1P  +\sigma^2}{h_1(P-p_0)+\sigma^2}\right)
+\sum_{k=1}^K B\log_2 \left(\frac{h_k(P-p_0)+\sigma^2}{h_k(P-p_0-p_k)+\sigma^2}\right),\tag{\theequation}\\
\textrm{s.t.}\quad \:
&\sum_{k=1}^Ka_k \leq B\log_2 \left(\frac{h_1P  +\sigma^2}{h_1(P-p_0)+\sigma^2}\right),\\
&a_k+B\log_2 \left(\frac{h_k(P-p_0)+\sigma^2}{h_k(P-p_0-p_k)+\sigma^2}\right)\geq R_k,
 \forall k \in \mathcal K,\\
&\sum_{k=0}^Kp_k = P,\\
&p_0\geq \frac P 2  + \frac{\theta +\sigma^2}{2h_1},\\
&p_k\geq 0, 0\leq a_k\leq R_k,\quad \forall k \in \mathcal K.
\end{align}
\end{subequations}
\end{lemma}

\itshape {Proof:}  \upshape
See Appendix B.
\hfill $\Box$

Note that the maximum rate limitation $0\leq a_k\leq R_k$ is added in constraint (\ref{al2min1_2}e), which will prove to be helpful in obtaining the optimal private message transmission power in closed form.

\subsection{Optimal Private Message Transmission Power}
Given the simplified problem in (\ref{al2min1_2}), next, we find the optimal private message transmission power.
Given rate allocation $\boldsymbol a$ and common message power control $p_0$, problem (\ref{al2min1_2}) becomes
\begin{subequations}\label{al2min2}
\begin{align}
\mathop{\max}_{\bar{\boldsymbol  p} }\quad&
\sum_{k=1}^K B\log_2\left(\frac{h_k(P-p_0)+\sigma^2}{h_k(P-p_0-p_k)+\sigma^2}\right),\tag{\theequation}\\
\textrm{s.t.}\quad \:
&\sum_{k=1}^Kp_k = P-p_0,\\
&p_k\geq p_k^{\min},\quad \forall k \in \mathcal K,
\end{align}
\end{subequations}
where $\bar{\boldsymbol p}=[p_1,p_2,\cdots,p_K]$ is a vector of power that is allocated to each user for receiving private message and
\begin{equation}\label{al2eq1}
p_k^{\min} = \left(1-2^{\frac{a_k-R_k}{B}}
\right)
\left( P-p_0+\frac{\sigma^2}{h_k}
\right).
\end{equation}
Due to constraint (\ref{al2min1_2}e), $p_k^{\min}$ is always non-negative.

Note that the fist term  in objective function (\ref{al2min1_2}) is a constant with given common message power control $p_0$, thus the  fist term  in objective function (\ref{al2min1_2}) is omitted in problem \eqref{al2min2}.
In (\ref{al2min2}b), $p_k^{\min}$  is used to meet the minimum rate constraint in (\ref{al2min1_2}b), and problem  \eqref{al2min2} is feasible if and only if $\sum_{k=1}^K p_k^{\min}\leq P-p_0$, which can be given as:
\begin{equation}\label{al2eq1_2}
\sum_{k=1}^K\left(1-2^{\frac{a_k-R_k}{B}}
\right)
\left( P-p_0+\frac{\sigma^2}{h_k}
\right) \leq P-p_0.
\end{equation}
Since objective function is convex, we can infer that the maximization
problem  (\ref{al2min2}) is nonconvex.
To effectively solve problem (\ref{al2min2}), the following theorem is presented.
\begin{theorem}
For the optimal solution $\bar{ \boldsymbol p}^*$ of problem (\ref{al2min2}), there exists one $k$ such that
$p_k^*=P-p_0-\sum_{j=1,j\neq k}^Kp_j^{\min}$ and $p_j^*=p_j^{\min}$, $\forall j \in \mathcal K, j \neq k$.
\end{theorem}

\itshape {Proof:}  \upshape
See Appendix C.
\hfill $\blacksquare$

From Theorem 1, the structure of the optimal solution of problem (\ref{al2min2}) is revealed.
Although problem  (\ref{al2min2}) is nonconvex, the optimal solution can be obtained in closed form, which can be given by the following theorem.

\begin{theorem}
For nonconvex problem  (\ref{al2min2}), the optimal power allocation $\bar{\boldsymbol p}^*$ is
\begin{equation}\label{al2eq2}
p_k^*=P-p_0-\sum_{j=1,j\neq k}^K\left(1-2^{\frac{a_j-R_j}{B}}
\right)
\left( P-p_0+\frac{\sigma^2}{h_j}
\right),
\end{equation}
\begin{equation}\label{al2eq2_2}
p_j^*=\left(1-2^{\frac{a_j-R_j}{B}}
\right)
\left( P-p_0+\frac{\sigma^2}{h_j}
\right), \quad \forall j \in \mathcal K, j \neq k,
\end{equation}
and the optimal sum-rate of private message is
\begin{align}\label{al2eq3}
&B\log_2 \left(\frac{ P-p_0 +\frac{\sigma^2}{h_k}}
{\sum_{j=1,j\neq k}^K \left(1-2^{\frac{a_j-R_j}{B}}
\right)
\left( P-p_0+\frac{\sigma^2}{h_j}
\right)+\frac{\sigma^2}{h_k}}\right)
+ \sum_{j=1,j\neq k}^K  (R_j-a_j),
\end{align}
where
\begin{equation}\label{al2eq5}
k=\arg\min_{j\in\mathcal K} 2^{\frac{a_j-R_j}{B}}
\left( P-p_0+\frac{\sigma^2}{h_j}
\right).
\end{equation}
\end{theorem}

\itshape {Proof:}  \upshape
See Appendix D.
\hfill $\blacksquare$

Theorem 2 states that it is optimal for the BS to allocate more power to the user that can maximize the sum-rate while allocating the minimum transmit power that can meet the data rate requirement for all other users.

For the special case with $a_j=R_j$, $\forall j\in\mathcal K$, we can obtain $p_j^{\min}=0$ and  $k=\arg\min_{j\in\mathcal K} \frac{\sigma^2}{h_j}=K$ according to \eqref{al2eq5}, i.e., all the power should be allocated to the user with the highest channel gain.
This observation is trivial since allocating the maximum power to the user with the highest channel gain will always improve the rate. 

For the special case in which $\boldsymbol a=\boldsymbol 0$, i.e., the broadcast channel without SIC, we have the following corollary that follows from Theorem 2.
\begin{coro}
For the downlink nonconvex sum-rate maximization problem in broadcast channel, which is given by:
\begin{subequations}\label{al2min2_2}
\begin{align}
\mathop{\max}_{\boldsymbol  p }\quad&\sum_{k=1}^K{B \log_2 \left( 1+ \frac{h_kp_k}{h_k\sum_{j=1,j\neq k}^Kp_j+\sigma^2}\right)}, \tag{\theequation}\\
\textrm{s.t.}\quad \:
&{B \log_2 \left( 1+ \frac{h_kp_k}{h_k\sum_{j=1,j\neq k}^Kp_j+\sigma^2}\right)}\geq R_k,\:
\forall k \in \mathcal K,\\
&\sum_{k=1}^Kp_k\leq P,\\
& p_k\geq 0,\quad \forall k \in \mathcal K,
\end{align}
\end{subequations}
the maximization problem in \eqref{al2min2_2} is feasible if and only if
\begin{equation}
\sum_{j=1}^K\left(1-2^{\frac{-R_j}{B}}
\right)
\left( P-p_0+\frac{\sigma^2}{h_j}
\right) \leq P-p_0,
\end{equation}
and
 the optimal power allocation $\boldsymbol p^*$
is to meet the minimum rate requirements of all users except one user, i.e.,
\begin{equation}
p_k^*=P-\sum_{j=1,j\neq k}^K\left(1-2^{\frac{-R_j}{B}}
\right)
\left( P+\frac{\sigma^2}{h_j}
\right),
\end{equation}
\begin{equation}
p_j^*=\left(1-2^{\frac{-R_j}{B}}
\right)
\left( P+\frac{\sigma^2}{h_j}
\right), \quad \forall j \in \mathcal K, j \neq k,
\end{equation}
and the optimal objective value is
\begin{align}
&B\log_2 \left(\frac{ P +\frac{\sigma^2}{h_k}}
{\sum_{j=1,j\neq k}^K \left(1-2^{\frac{-R_j}{B}}
\right)
\left( P+\frac{\sigma^2}{h_j}
\right)+\frac{\sigma^2}{h_k}}\right)
+ \sum_{j=1,j\neq k}^K R_j,
\end{align}
where
\begin{equation}\label{al2min2_2eq1}
k=\arg\min_{j\in\mathcal K} 2^{\frac{-R_j}{B}}
\left( P+\frac{\sigma^2}{h_j}
\right).
\end{equation}
\end{coro}

Corollary 1 provides the optimal power allocation that maximizes the downlink sum-rate of the broadcast channel. From this result, we can see that more power should be allocated to one of the users compared to allocating the minimum transmit power to all other users.
The user that should be allocated more power is jointly determined by the channel gain and the minimum rate demand, as shown in \eqref{al2min2_2eq1}.

\subsection{Optimal Rate Allocation}
In the previous subsection, the optimal power allocation vector $\bar{\boldsymbol p}$ can be obtained as a function of the rate allocation vector $\boldsymbol a$ and common message power $p_0$.
Thus, substituting the optimal power allocation vector $\bar{\boldsymbol p}$ given in \eqref{al2eq2} and \eqref{al2eq2_2} in Theorem 2, the original problem in (\ref{al2min1_2}) can be simplified as:
\begin{subequations}\label{al2min3}
\begin{align}
\mathop{\max}_{\boldsymbol a, p_0 }\quad&   B\log_2 \left(\frac{h_1P  +\sigma^2}{h_1(P-p_0)+\sigma^2}\right)   + \sum_{j=1,j\neq k}^K  (R_j-a_j)
 \nonumber\\
&+B\log_2\left(\frac{ P-p_0 +\frac{\sigma^2}{h_k}}
{\sum_{j=1,j\neq k}^K \left(1-2^{\frac{a_j-R_j}{B}}
\right)
\left( P-p_0+\frac{\sigma^2}{h_j}
\right)+\frac{\sigma^2}{h_k}}\right),\tag{\theequation}
\\
\textrm{s.t.}\quad \:
&\sum_{j=1}^Ka_j \leq B\log_2 \left(\frac{h_1P  +\sigma^2}{h_1(P-p_0)+\sigma^2}\right), \\
&k=\arg\min_{j\in\mathcal K} 2^{\frac{a_j-R_j}{B}}
\left( P-p_0+\frac{\sigma^2}{h_j}
\right),\\
&\sum_{j=1}^K
\left(1-2^{\frac{a_j-R_j}{B}}
\right)
\left( P-p_0+\frac{\sigma^2}{h_j}
\right) \leq P-p_0,\\
&p_0\geq \frac P 2  + \frac{\theta +\sigma^2}{2h_1},\\
&0\leq a_j\leq R_j,\quad \forall j \in \mathcal K,
\end{align}
\end{subequations}
where constraint (\ref{al2min3}b) is added since more power should be allocated to user $k$ to maximize the sum-rate while other users are allocated the minimum power to ensure the minimum rate constraint.
Constraint (\ref{al2min3}c) follows from \eqref{al2eq1_2}, which ensures that the private message power control problem is feasible.

Due to objective function  and constraints (\ref{al2min3}a)-(\ref{al2min3}c), problem (\ref{al2min3}) is nonconvex and, hence, it is generally hard to directly optimize rate allocation  $\boldsymbol a$ and private power $p_0$.
To solve problem (\ref{al2min3}), we first fix the private message transmission power and derive the optimal rate allocation.
Given common message transmission power $p_0$, problem (\ref{al2min3}) becomes
\begin{subequations}\label{al2min3_2}
\begin{align}
\mathop{\max}_{\boldsymbol a  }\quad&  \sum_{j=1,j\neq k}^K  (R_j-a_j)
 +   B\log_2  \left(\frac{ P-p_0 +\frac{\sigma^2}{h_k}}
{\sum_{j=1,j\neq k}^K \left(1-2^{\frac{a_j-R_j}{B}}
\right)
\left( P-p_0+\frac{\sigma^2}{h_j}
\right)+\frac{\sigma^2}{h_k}}\right),  \tag{\theequation} \\
\textrm{s.t.}\quad \:
&\sum_{j=1}^Ka_j \leq c_1,\\
&k=\arg\min_{j\in\mathcal K} 2^{\frac{a_j-R_j}{B}}
\left( P-p_0+\frac{\sigma^2}{h_j}
\right),\\
&\sum_{j=1}^K
\left(1-2^{\frac{a_j-R_j}{B}}
\right)
\left( P-p_0+\frac{\sigma^2}{h_j}
\right) \leq P-p_0,\\
& 0\leq a_j\leq R_j,\quad \forall j \in \mathcal K.
\end{align}
\end{subequations}

To solve \eqref{al2min3_2}, we first need to prove that the objective  function (\ref{al2min3_2}) is convex, which can be given by the following lemma.
\begin{lemma}
The objective function (\ref{al2min3_2}) is convex.
\end{lemma}

\itshape {Proof:}  \upshape
See Appendix E.
\hfill $\blacksquare$

According to Lemma 4, the objective  function (\ref{al2min3_2}) is convex.
The maximum point of the convex function always lies in the corner points, i.e., the optimal solution  of problem (\ref{al2min3_2}) lies in a finite space of potential candidates.
For the optimal solution of (\ref{al2min3_2}), we can prove the following:
\begin{theorem}
The optimal solution $\boldsymbol a^*$ of problem (\ref{al2min3_2})  exists in the following three cases:

Case 1:
\begin{equation}
a_j^*\in\{0,R_j\}, \quad j \in \mathcal K.
\end{equation}

Case 2: there exists $l\in\mathcal K$ such that
\begin{equation}
a_j^*\in\{0,R_j\},a_l^*\in\{b_1|_0^{R_k},b_2|_0^{R_k}\}\quad  j \in \mathcal K, j\neq l,
\end{equation}
where
\begin{align}
b_1=& c_1-\sum_{j=1,j\neq k}^Ka_j,
\end{align}
and
\begin{align}
b_2=&R_k-B\log_2\left( P-p_0+\frac{\sigma^2}{h_k}
\right)
+B \log_2\left({\sum_{j=1,j\neq k}^K \left(1-2^{\frac{a_j-R_j}{B}}
\right)
\left( P-p_0+\frac{\sigma^2}{h_j}
\right)+\frac{\sigma^2}{h_k}}\right).
\end{align}


Case 3: there exists $m,n\in\mathcal K$,
\begin{equation}
a_j^*\in\{0,R_j\}, a_m^*=c_1-\sum_{j=1,j\neq k}^Ka_j^*,\quad  j \in \mathcal K, j\neq m,n,
\end{equation}
and $a_n^*$ satisfies
\begin{align}\label{al2th3eq1}
&\left(1-2^{\frac{c_1-\sum_{j=1,j\neq k,n}^Ka_j^*-a_n^*-R_m}{B}}
\right)\left( P-p_0+\frac{\sigma^2}{h_m}
\right)
 +\left(1-2^{\frac{a_n^*-R_n}{B}}
\right)
\left( P-p_0+\frac{\sigma^2}{h_n}
\right)
\nonumber\\
&+\sum_{j=1,j\neq m,n}^K
\left(1-2^{\frac{a_j^*-R_j}{B}}
\right)
\left( P-p_0+\frac{\sigma^2}{h_j}
\right)= P-p_0.
\end{align}
\end{theorem}


In Theorem 3, Case 1 indicates that the corner points satisfy the constraint in (\ref{al2min3_2}d),  Case 2 indicates that the corner points satisfy the constraint in (\ref{al2min3_2}d) as well as one constraint in (\ref{al2min3_2}a) or (\ref{al2min3_2}c), and  Case 3 indicates that the corner points satisfy the constraints in (\ref{al2min3_2}a), (\ref{al2min3_2}c), and (\ref{al2min3_2}d).
Theorem 3 follows directly from the fact that maximizing a convex function lies in its corner points.
The optimal solution of problem (\ref{al2min3_2}) will be one of the corner points in all three cases.

Since the left term of \eqref{al2th3eq1} is a convex function of $a_n^*$, at most two solutions for $a_n^*$ satisfy \eqref{al2th3eq1}.
We then observe that there are $2^K$, $2^K$, and $2^{K-1}$ points for Cases 1, 2 and 3, respectively.
As a result, there are $5\times2^{K-1}$ potential candidates for the optimal solution of problem (\ref{al2min3_2}) according to Theorem 3.

For the special case that all users have the same data rate requirement, the optimal solution of problem (\ref{al2min3_2}) can be precisely formulated in the following theorem.
\begin{theorem}
Given the data rate requirement of each user, i.e., $R_1=R_2=\cdots=R_K=R$,
 the optimal solution $\boldsymbol a^*$ of problem (\ref{al2min3_2}) is:

 i) if $\lfloor \frac{c_1}{R} \rfloor<K$,
\begin{equation}\label{al2th4eq1}
r_{j}^*=\left\{ \begin{array}{ll}
\!\!R, &\text{if}\; j<l,\\
\!\!c_1 -(l-1)R, &\text{if}\; j=l,\\
\!\!0, &\text{otherwise,}
\end{array} \right.
\end{equation}
where $\lfloor \cdot \rfloor$ means round down and $l=\lfloor  \frac{c_1}{R} \rfloor$.

ii) if $\lfloor \frac{c_1}{R} \rfloor\geq K$,
\begin{equation}\label{al2th4eq2}
r_{j}^*=R,  \quad \forall j \in\mathcal K.
\end{equation}
\end{theorem}

\itshape {Proof:}  \upshape
See Appendix F.
\hfill $\blacksquare$

Theorem 4 provides the optimal rate allocation of problem (\ref{al2min3_2}) with equal rate demand in closed form.
Based on Theorem 4, we can determine the user that the BS will allocate additional power using the following lemma.
\begin{lemma}
Given the data rate requirement of each user, i.e., $R_1=R_2=\cdots=R_K=R$, it is optimal to allocate the additional power to the user with the highest channel gain, which is given by:
\begin{equation}
k=\arg\min_{j\in\mathcal K} 2^{\frac{a_j-R}{B}}
\left( P-p_0+\frac{\sigma^2}{h_j}
\right)=K.
\end{equation}
\end{lemma}

\itshape {Proof:}  \upshape
Based on Theorem 5, we have $a_1^*\geq a_2^*\geq\cdots\geq a_K^*$.
Since $\frac{\sigma^2}{h_1}\geq\frac{\sigma^2}{h_2}\geq\cdots\geq\frac{\sigma^2}{h_K}$, we have
\begin{align}
k=\arg\min_{j\in\mathcal K} 2^{\frac{a_j^*-R}{B}}
\left( P-p_0+\frac{\sigma^2}{h_j}
\right)=K.
\end{align}
\hfill $\blacksquare$


%
%
%
%
%

\begin{lemma}
For a network with two users, $K=2$ and their data rate requirements are $R_1=R_2=R$, the optimal solution $(\boldsymbol a^*,p_0^*)$ of problem (\ref{al2min3})
is $a_1^*=R$, $a_2^*=0$
and
\begin{equation}
p_0^* =\max\left\{\left(1-2^{\frac{-R}{B}}
\right)
\left( P+\frac{\sigma^2}{h_1}
\right),\frac P 2  + \frac{\theta +\sigma^2}{2h_1} \right\}.
\end{equation}
\end{lemma}

\itshape {Proof:}  \upshape
See Appendix G.
\hfill $\blacksquare$

According to Lemma 6, for a network that only has two users, it is optimal to decode the messages of user 1 first in the common message, and the remaining power is then all allocated to user 2.
%
For $K=2$ users, NOMA can be viewed as a special case of RSMA if $a_2=p_1=0$, i.e., the common message is allocated to only user 1, while only user 2 has its unique private message.
\subsection{Optimal Rate Allocation and Power Control}

To obtain the optimal rate allocation and power control of problem (\ref{al2min1_2}), we propose a novel solution, shown in Algorithm 1 where $\xi$ is the minimum step size for searching $p_0$.
In Algorithm~1, the optimal common message power $p_0$ is obtained by a one-dimensional search method, while the optimal rate allocation $\boldsymbol a$ and private message power $\bar{\boldsymbol p}$ are accordingly obtained in closed form given $p_0$.

\begin{algorithm}[h]
\caption{Optimal Rate Allocation and Power Control}
\begin{algorithmic}[1]
\FOR{$p_0=\frac P 2  +\frac{\theta}{2} +\frac{\sigma^2}{2h_1} :\xi:P$}
\STATE Obtain the optimal rate allocation $\boldsymbol a$ of problem \eqref{al2min3_2}  according to Theorem 3.
\STATE Calculate the optimal common message power allocation $\bar{\boldsymbol p}$ of  (\ref{al2min2}) according to Theorem~2.
\ENDFOR
\STATE Obtain the optimal $p_0$ with the maximum objective value (\ref{al2min1_2}a).
\end{algorithmic}
\end{algorithm}

In each step of Algorithm 1, the main complexity lies in solving  \eqref{al2min3_2} given $p_0$.
Since there are $5\times2^{K-1}$ potential candidates for the optimal solution of problem (\ref{al2min3_2}) according to Theorem 3, the complexity of solving the problem in \eqref{al2min3_2} is $\mathcal O(5\times2^{K-1})$.
As a result, the complexity of Algorithm 1 is  $\mathcal O(5L 2^{K-1})$, where $L=\mathcal O\left(\frac{  P h_1  -  \theta h_1 - {\sigma^2} } {2h_1\xi}\right)$ denotes the number of iterations for searching $p_0$.
In practice, we consider a small number of users, i.e., $K$ is small due to the complexity of decoding the common message, the computation complexity of Algorithm 1 can be practical. {\color{myc1}{To deal with a large number of users, the users can be classified into different groups with small number of users in each group. The users in different groups occupy different subchannels and users in the same group are allocated to the same subchannel using RSMA.}}
 Moreover, for equal rate demand, the complexity  of solving the problem in \eqref{al2min3_2} is $\mathcal O(K)$ according to Theorem 4 and the complexity of Algorithm 1 is  $\mathcal O(LK)$.

\section{Simulation Results}
In this section, we evaluate the performance of the proposed optimal rate allocation and power control algorithm.
$K$ users there are uniformly distributed in a square area of size $300$ m $\times$ $300$~m.
The path loss model is $128.1+37.6\log_{10} d$ ($d$ is in km) \cite{8352643}
and the standard deviation of shadow fading is $4$ dB.
In addition, the bandwidth of the BS is $B=1$ MHz and the noise power is  $\sigma^2=-104$ dBm.
Unless specified otherwise, the system parameters are set as  maximum transmit power $P=30$ dBm, equal rate demand $R_1=R_2=\cdots=R_K=R=1$ Mbits/s and SIC detection threshold is set as $\theta=-94$ dBm.
The main system parameters are listed in Table~I.

\begin{table}[t]
\centering
\caption{System  Parameters} \label{tab:complexity}
\begin{tabular}{ccc}
  \hline
  \hline
  Parameter &   Value \\ \hline
Bandwidth of the BS $B$ &1 MHz  \\
noise power  $\sigma^2$& -104 dBm \\
maximum transmit power $P$ & 30 dBm\\
Minimal rate demand $R$& 1 Mbits/s\\
SIC detection threshold $\theta$& -94 dBm\\
  \hline
  \hline
\end{tabular}
\end{table}

The proposed optimal rate allocation and power control algorithm for rate maximization of RSMA is labeled as `RSMA'.
We compare with the proposed algorithm with the optimal power control of NOMA for rate maximization \cite{8352643}, which is labeled as `NOMA'.
To compare conventional orthogonal multiple access (OMA), we use a OFDMA system \cite{seong2006optimal} as a baseline, which is labeled as `OFDMA'. 

\begin{figure}[t]
\centering
\includegraphics[width=5in]{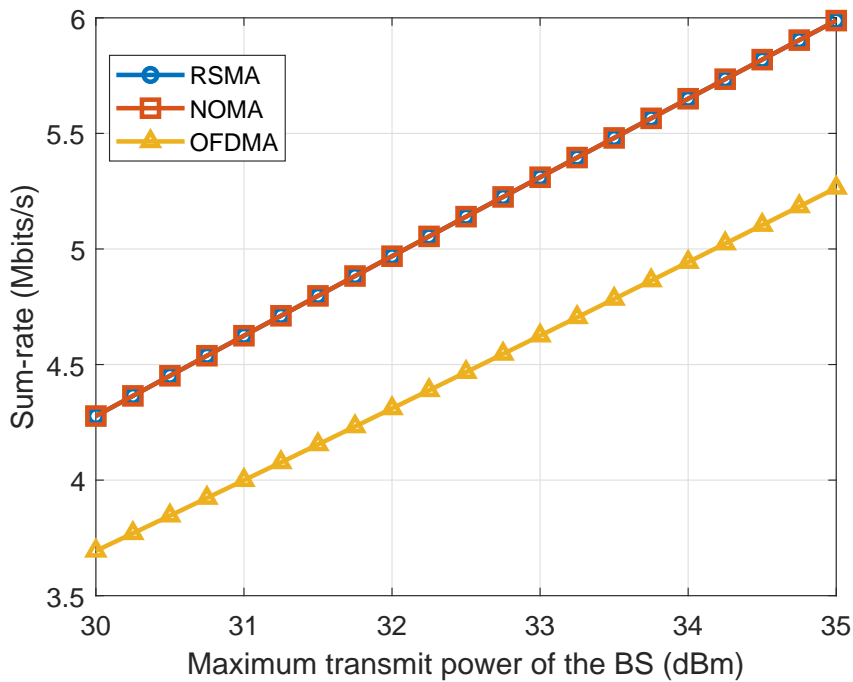}
\caption{Sum-rate versus maximum transmit power of the BS ($K=2$ users)} \label{fig2}
\end{figure}

\begin{figure}[htpb]
\centering
\includegraphics[width=5in]{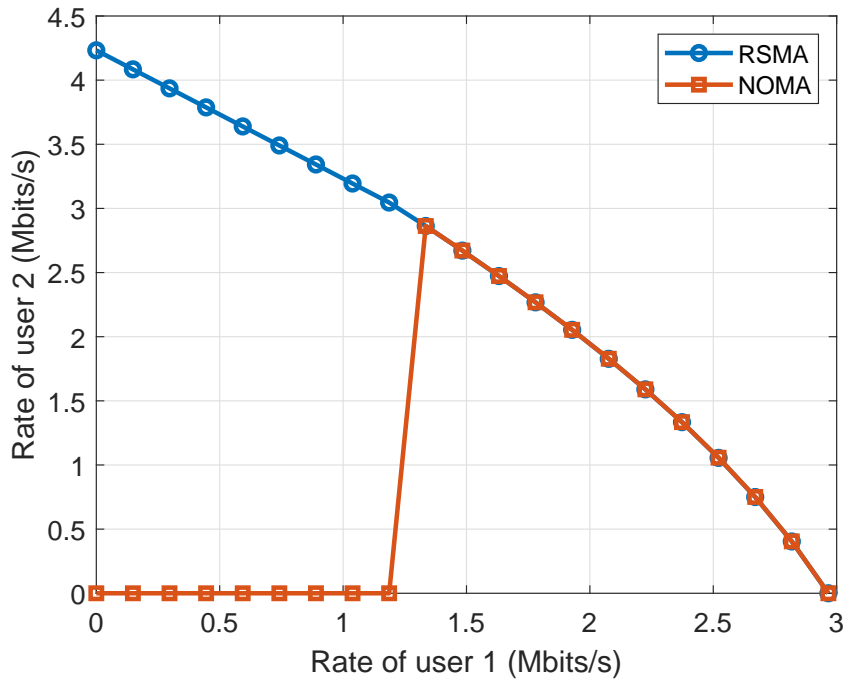}
\caption{Rate region for two users.} \label{fig9}
\end{figure}

Fig. \ref{fig2} shows how the sum-rate changes as the maximum transmit power of the BS varies for a network having two users. From Fig. \ref{fig2}, we can see that the sum-rate linearly increases with the logarithmic maximum transmit power of the BS.
This figure shows that RSMA achieves the same performance as NOMA in terms of sum-rate, which verifies the theoretical finding in Lemma 6.
{\color{myc1}{However, RSMA will dynamically allocate the rate of the common message to multiple users to meet the rate demand, while the rate decoded in NOMA is allocated to only one specific user.}}
From Fig. \ref{fig2}, we can see that both RSMA and NOMA significantly outperform conventional OFDMA.
This is because the BS can simultaneously transmit signals to all users by RSMA or NOMA at the same frequency, while the BS transmits signal to different users in different resource blocks by OFDMA.

\begin{figure}[htpb]
\centering
\includegraphics[width=5in]{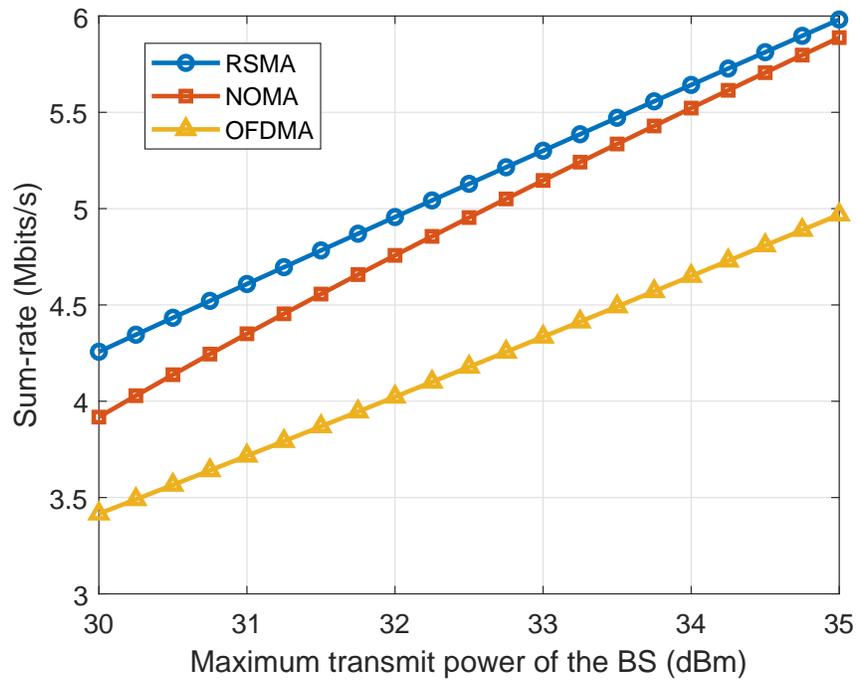}
\caption{Sum-rate versus maximum transmit power of the BS ($K=3$ users and $R_1=R_2=R_3=1$ Mbits/s).} \label{fig3}
\end{figure}

\begin{figure}[htpb]
\centering
\includegraphics[width=5in]{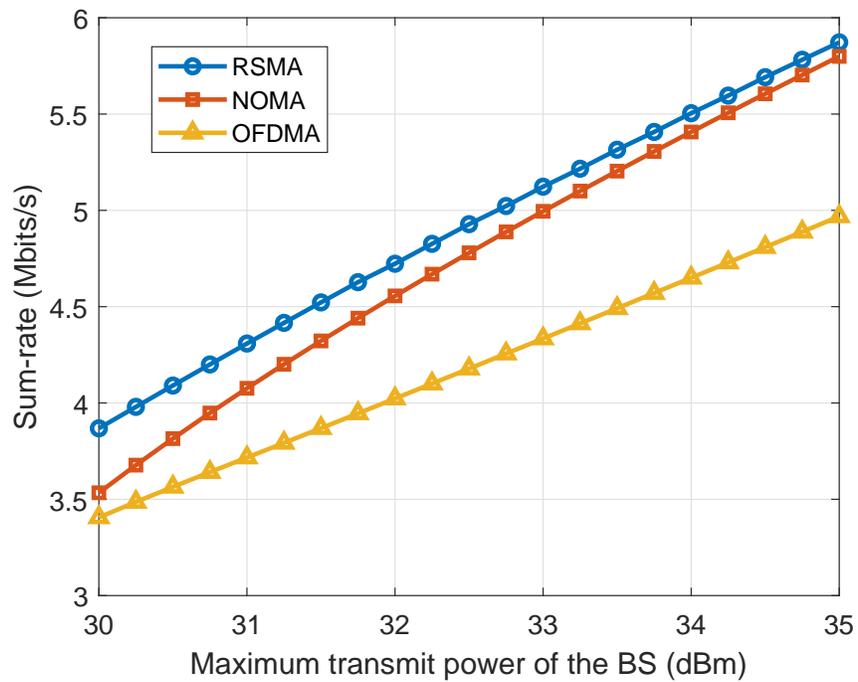}
\caption{Sum-rate versus maximum transmit power of the BS ($K=3$ users, $R_1=1.5$ Mbits/s, $R_2=0.5$ Mbits/s and $R_3=1$ Mbits/s).} \label{fig3_2}
\end{figure}

The rate region for two users is given in Fig.~\ref{fig9}.
It is shown that the rate of user 1 is always greater than a fixed value when the rate of users 2 is positive in NOMA. This is due to the fact that the allocated power of user 1 should be greater than a fixed value according to the successful SIC constraint.
The maximum rate of user 2 in RSMA is larger than that in NOMA due to the benefit of rate splitting in RSMA.

Figs.~\ref{fig3} and \ref{fig3_2} show the sum-rate sum-rate versus maximum transmit power of the BS with  equal rate demand and unequal rate demand for three users, respectively.
From these figures, we can see that RSMA always achieves the best performance among all schemes.
This is due to the fact that the number of SIC in RSMA is once, while the number of SIC in NOMA can be twice, which results in high power allocation to the users with low channel gain according to the successful SIC power requirement \cite[Eq. (3)]{8352643} and leads to a low sum-rate.
These figures also show that RSMA outperforms NOMA particularly for low maximum BS transmit power.
Compared to OFDMA, RSMA is better due to the fact that all users can be served with the whole bandwidth of the BS.

\begin{figure}[t]
\centering
\includegraphics[width=5in]{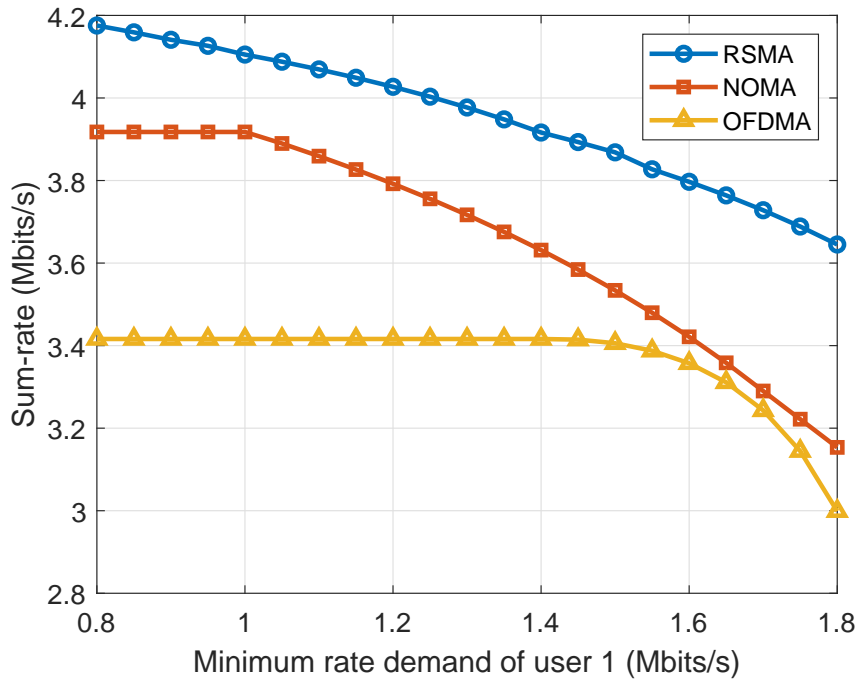}
\caption{Sum-rate versus minimum rate demand ($K=3$ users, $R_2=R_3=1$ M bits/s).} \label{fig5}
\end{figure}

\begin{figure}[htpb]
\centering
\includegraphics[width=5in]{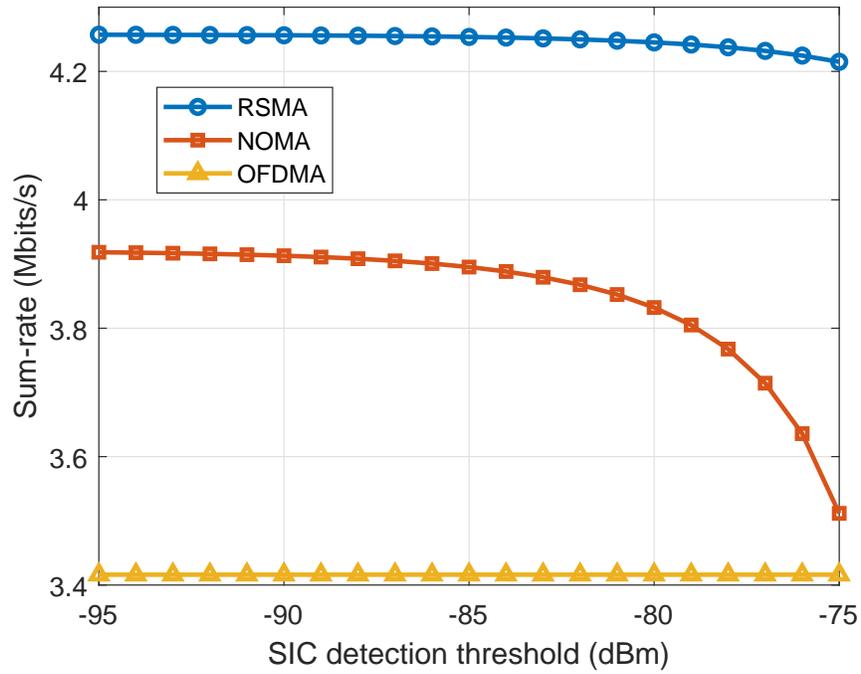}
\caption{Sum-rate versus SIC detection threshold ($K=3$ users).} \label{fig7}
\end{figure}

\begin{figure}[htpb]
\centering
\includegraphics[width=5in]{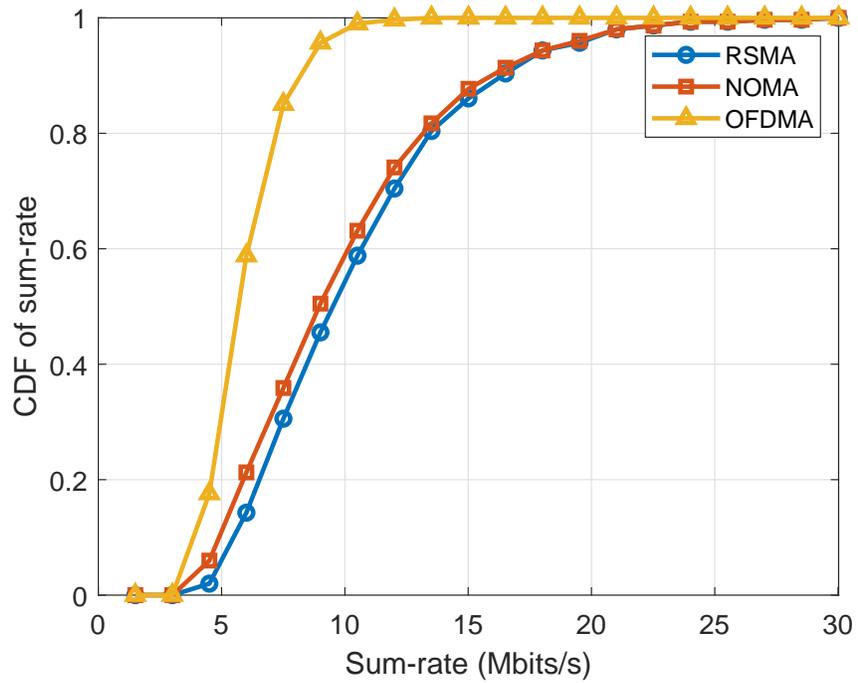}
\caption{CDF of the sum-rate resulting from RSMA, NOMA, and OFDMA for a network with $K=3$ users.} \label{fig8}
\end{figure}

\begin{figure}[htpb]
\centering
\includegraphics[width=5in]{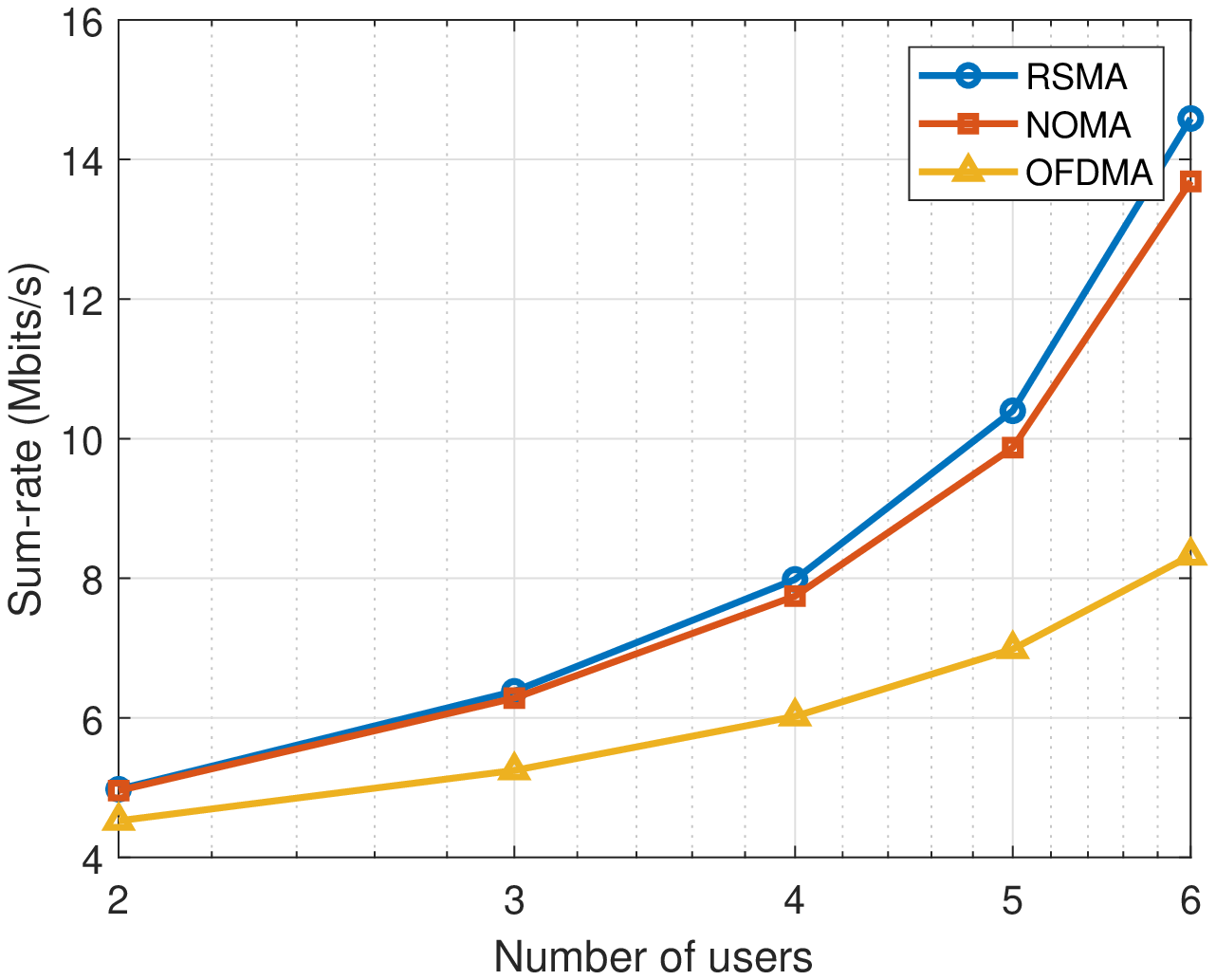}
\caption{Sum-rate versus number of users.} \label{fig6}
\end{figure}

Fig. \ref{fig5} shows the sum-rate versus minimum rate demand.
From this figure, RSMA always achieves a better performance than NOMA and OFDMA.
From Fig. \ref{fig5}, we can observe that the sum-rate decreases slightly when minimum rate demand is low.
However, for a high minimum rate demand, the sum-rate decreases rapidly.
This is because a high minimum rate demand requires the BS to allocate more power to the users with worse channel gains, which consequently degrades the sum-rate performance.
Fig. 6 also demonstrates that, as the minimum rate demand increases, the sum-rates of OFDMA and NOMA decrease faster than RSMA.
In particular, RSMA can achieve up to 15.6\% and 21.5\% gains in terms of data rate compared to NOMA and OFDMA, respectively.
This is due to the fact that RSMA exhibits a better spectrum efficiency compared to OFDMA and NOMA, and OFDMA and NOMA are more sensitive to high minimum rate demand than RSMA.

Fig. \ref{fig7} shows the sum-rate versus SIC detection threshold $\theta$.
For both RSMA and NOMA, we find
that the sum-rate decreases as the SIC detection threshold increases. This is
due to the fact that, as the SIC detection threshold increases, the BS must allocate more power to the common message in RSMA and the user with worse channel gain in NOMA. For OFDMA, naturally, the sum-rate remains the
same when the SIC detection threshold increases. 
The proposed RSMA algorithm outperforms the NOMA in terms
of sum-rate, particularly for cases with a high SIC detection threshold.
 Moreover, the sum-rate decreases faster for NOMA than RSMA as the SIC detection threshold increases,
 which implies that RSMA is more suitable for high SIC detection threshold.

Fig. \ref{fig8} shows the cumulative distribution function (CDF) of the sum-rate resulting from RSMA, NOMA, and OFDMA for a network with $K=3$ users.
From Fig. \ref{fig8}, the CDFs for
RSMA and NOMA all improve significantly vs.
OFDMA especially for high sum-rate region, which shows that both RSMA and NOMA are suitable for high sum-rate transmission.
Moreover, we can find that RSMA outperforms NOMA at regions with moderate data rates, i.e., 5-15 Mbits/s. This is because RSMA can adjust the split between the common and private messages so as to control the interference decoding and thus optimize the sum-rate of users.

{\color{myc1}{The sum-rate versus number of users is given in Fig.~\ref{fig6}.
Clearly, the proposed RSMA is always better than NOMA and OFDMA especially when the number of users is large. When the number
of users is large, the multiuser gain is more apparent by the proposed RSMA compared to conventional NOMA and OFDMA. This is because RSMA can effectively determine the rate of each user receiving common message to meet its specific rate demand, while the SIC time of each user is high for NOMA and the allocated bandwidth of each user is low for OFDMA when the number of users is large. RSMA achieves better performance than NOMA and OFDMA at the cost of additional computational complexity according to Section III-D.
}}

\section{Conclusions}

In this paper, we have investigated the allocation of data rate of common message transmission and the transmit power used for common and private message transmission in a SISO RSMA system. We have formulated the problem as a minimization problem. To solve this problem, we have derived the optimal transmit power of private message in closed form.
Then, we have characterized the finite solution space for the optimal rate allocation.
Finally, we have proposed a one-dimensional search algorithm to find the optimal rate allocation and power control solutions.
Simulation results show that RSMA achieves higher sum-rate than NOMA and OFDMA especially for low maximum transmit power of the BS, high minimum rate demand of users, high SIC detection threshold and large number of users.

\appendices
\section{Proof of Lemma 2}
\setcounter{equation}{0}
\renewcommand{\theequation}{\thesection.\arabic{equation}}

We assume that the optimal solution $(\boldsymbol a^*, \boldsymbol p^*)$ of problem (\ref{sys1min1}) satisfies $\sum_{k=0}^Kp_k^*<P$.
We construct new power allocation by scaling power $\boldsymbol p^*$, i.e.,
\begin{equation}\label{AppLema2eq1}
p_k'=\frac{P}{\sum_{j=0}^Kp_j^*}p_k^*>p_k^*, \quad \forall k=0,1,\cdots,K.
\end{equation}
Given new power allocation vector $\boldsymbol p'=[p_0',p_1',\cdots,p_K']$, we have
\begin{align}\label{AppLema2eq2}
& B \log_2 \left( 1+ \frac{h_1p_0'}{h_1\sum_{j=1}^Kp_j'+\sigma^2}
\right)
>  B \log_2 \left( 1+ \frac{h_1p_0^*}{h_1\sum_{j=1}^Kp_j^*+\frac{\sigma^2{\sum_{j=0}^Kp_j^*}}
{ {P}}}
\right),
\end{align}
\begin{align}\label{AppLema2eq2_2}
& B \log_2 \left( 1+ \frac{h_kp_k'}{h_k\sum_{j=1,j\neq k}^Kp_k'+\sigma^2}
\right)
>  B \log_2 \left( 1+ \frac{h_kp_k^*}{h_1\sum_{j=1,j\neq k}^Kp_j^*+\frac{\sigma^2{\sum_{j=0}^Kp_j^*}}
{ {P}}}
\right),
\end{align}
and
\begin{align}\label{AppLema2eq3}
 p_0' -\sum_{j=1}^K p_j'=&\frac{P}{\sum_{j=0}^Kp_j^*}\left(p_0^* -\sum_{j=1}^K p_j^* \right)
> p_0^* -\sum_{j=1}^K p_j^*\geq \theta+\frac{\sigma^2}{h_1} ,
\end{align}
where the first inequalities in \eqref{AppLema2eq2}-\eqref{AppLema2eq3} follow from the fact that $\frac{P}{\sum_{j=0}^Kp_j^*}>1$.

According to \eqref{AppLema2eq2}-\eqref{AppLema2eq3}, we can see that new solution $(\boldsymbol a^*, \boldsymbol p')$ is feasible and the objective value (\ref{sys1min1}) of new solution is better than that of solution $(\boldsymbol a^*, \boldsymbol p^*)$, which contradicts the fact that  $(\boldsymbol a^*, \boldsymbol p^*)$ is the optimal solution.
Lemma 2 is proved.

\section{Proof of lemma 3}
\setcounter{equation}{0}
\renewcommand{\theequation}{\thesection.\arabic{equation}}

If the pair $(\boldsymbol a, {\boldsymbol p})$ is feasible in problem (\ref{al2min1_2}), then the pair $(\boldsymbol a, {\boldsymbol p})$ is also feasible in problem (\ref{al2min1}) with the same objective value. It follows from the fact that the optimal value of (\ref{al2min1_2})
is less than or equal to the optimal value of (\ref{al2min1}).

Conversely, if the pair $(\boldsymbol a, {\boldsymbol p})$ is feasible in (\ref{al2min1}),
we can construct a new pair $(\boldsymbol a', {\boldsymbol p})$,
where
\begin{equation}
a_k'=\min\{ a_k,R_k\}, \quad \forall k\in \mathcal K.
\end{equation}
It can be shown that solution $(\boldsymbol a', {\boldsymbol p})$ is feasible in problem (\ref{al2min1_2}).
Moreover, the objective value of problem (\ref{al2min1}) is the same as problem (\ref{al2min1_2}).
Thus, we conclude that the
optimal value of (\ref{al2min1_2}) is greater than or equal to the optimal value of (\ref{al2min1}). Hence, problem (\ref{al2min1_2}) is equivalent to problem (\ref{al2min1}).

%

\section{Proof of Theorem 1}
\setcounter{equation}{0}
\renewcommand{\theequation}{\thesection.\arabic{equation}}

Assume that there exist $m$ and $n$ such that $p_m^*>p_m^{\min}$ and $p_n^*>p_n^{\min}$ for the optimal solution $\bar { \boldsymbol p}^*$.
Next, we can show that there always exist feasible power $p_m'$ and $p_n'$ with better objective value (\ref{al2min2}a).

To construct such $p_m'$ and $p_n'$, we substitute $p_j=p_j^*$, $j\in\mathcal K$, $j\neq m,n$, into problem  (\ref{al2min2}), which yields
\begin{subequations}\label{AppTh1min1}
\begin{align}
\mathop{\max}_{p_m, p_n }\quad&
B\log_2 \left(\frac{h_m(P-p_0)+\sigma^2}{h_m(P-p_0-p_m)+\sigma^2}\right) +B\log_2 \left(\frac{h_n(P-p_0)+\sigma^2}{h_n(P-p_0-p_n)+\sigma^2}\right),\tag{\theequation}\\
\textrm{s.t.}\quad \:
&p_m+p_n= P-p_0-\sum_{j=1,j\neq m, n}^K p_j^*,\\
&p_m\geq p_m^{\min},p_n\geq p_n^{\min}.
\end{align}
\end{subequations}
According to (\ref{AppTh1min1}a), we have
\begin{equation}\label{AppTh1eq1}
p_m=P-p_0-\sum_{j=1,j\neq m, n}^K p_j^*-p_n.
\end{equation}
Combining (\ref{AppTh1min1}b) and \eqref{AppTh1eq1}, we have
\begin{equation}\label{AppTh1eq2}
p_n^{\min}\leq p_n\leq P-p_0-\sum_{j=1,j\neq m, n}^K p_j^*-p_m^{\min}.
\end{equation}

From \eqref{AppTh1eq1} and \eqref{AppTh1eq2}, we can see that problem (\ref{AppTh1min1}) can be simplified as
\begin{subequations}\label{AppTh1min2}
\begin{align}
\mathop{\max}_{ p_n }\quad&
-B\log_2 \left({h_m\left(\sum_{j=1,j\neq m, n}^K p_j^*+p_n\right)+\sigma^2}\right) - B\log_2 ({h_n(P-p_0-p_n)+\sigma^2}),\tag{\theequation}\\
\textrm{s.t.}\quad \:
&p_n^{\min}\leq p_n\leq P-p_0-\sum_{j=1,j\neq m, n}^K p_j^*-p_m^{\min}.
\end{align}
\end{subequations}

Due to the convexity of function $-\log(x)$, the objective function (\ref{AppTh1min2}) is convex.
Since the maximization of a convex function always lies in the boundary of the feasible solution, i.e., the optimal solution $p_n'$ of problem  (\ref{AppTh1min2}) satisfies
\begin{equation}\label{AppTh1eq3}
p_n'\in\left\{p_n^{\min}, P-p_0-\sum_{j=1,j\neq m, n}^K p_j^*-p_m^{\min}\right\}.
\end{equation}
Further considering \eqref{AppTh1eq1}, we can construct
\begin{equation}\label{AppTh1eq5}
p_m'=P-p_0-\sum_{j=1,j\neq m, n}^K p_j^*-p_n'.
\end{equation}

Based on the equivalence of problem (\ref{AppTh1min1}) and problem (\ref{AppTh1min2}),
$(p_m',p_n')$ is the optimal solution of problem (\ref{AppTh1min1}).
According to \eqref{AppTh1eq3} and \eqref{AppTh1eq5},
$p_m'=p_m^{\min}$ or $p_n'=p_n^{\min}$ is always satisfied.

Since $(p_m',p_n')$ is the optimal solution of problem (\ref{AppTh1min1}) and $(p_m',p_n') \neq (p_m^*,p_n^*)$,
we can claim that
solution
\begin{equation}
(p_1^*,\cdots, p_{m-1}^*,p_m',p_{m+1}^*,\cdots, p_{n-1}^*, p_n',p_{n+1}^*,\cdots, p_{K}^*)
\end{equation}
is feasible with better objective value than solution $\bar {\boldsymbol p}^*$, which contradicts the fact that $\bar{ \boldsymbol p}^*$ is the optimal solution of problem (\ref{al2min2}).

As a result, the proof of Theorem 1 is complete.

\section{Proof of Theorem 2}
\setcounter{equation}{0}
\renewcommand{\theequation}{\thesection.\arabic{equation}}

Substituting the optimal power allocation $p_k^*=P-p_0-\sum_{j=1,j\neq k}^Kp_j^{\min}$ and $p_j^*=p_j^{\min}$, $\forall j \in \mathcal K, j \neq k$ to problem (\ref{al2min2}) according to Theorem 1, we can obtain the objective value (\ref{al2min2}a) as
\begin{align}\label{AppTh2eq1}
&\sum_{j=1,j\neq k}^K B\log_2\left(\frac{h_j(P-p_0)+\sigma^2}{h_j(P-p_0-p_j^{\min})+\sigma^2}\right)
+B\log_2\left(\frac{h_k(P-p_0)+\sigma^2}{h_k\sum_{j=1,j\neq k}^Kp_j^{\min}+\sigma^2}\right).
\end{align}

To maximize sum-rate \eqref{AppTh2eq1}, the optimal $k$ should be chosen as
\begin{align}\label{AppTh2eq2}
k=
\arg\max_{m\in\mathcal K}&\sum_{j=1}^K B\log_2\left(\frac{h_j(P-p_0)+\sigma^2}{h_j(P-p_0-p_j^{\min})+\sigma^2}\right)
 - B\log_2\left(\frac{h_m(P-p_0)+\sigma^2}{h_m(P-p_0-p_m^{\min})+\sigma^2}\right)
 \nonumber\\
 &+ B\log_2\left(\frac{h_m(P-p_0)+\sigma^2}{h_m\sum_{j=1,j\neq m}^Kp_j^{\min}+\sigma^2}\right)\nonumber\\
=\arg\max_{m\in\mathcal K}\:& B\log_2 ({h_m(P-p_0-p_m^{\min})+\sigma^2})
- B\log_2 \left({h_m\sum_{j=1,j\neq m}^Kp_j^{\min}+\sigma^2}\right)
\nonumber\\
=\arg\max_{m\in\mathcal K}&
\frac{h_m(P-p_0-p_m^{\min})+\sigma^2}
{h_m\sum_{j=1,j\neq m}^Kp_j^{\min}+\sigma^2}-1
\nonumber\\
=\arg\max_{m\in\mathcal K}&
\frac{P-p_0-\sum_{j=1}^Kp_j^{\min}}
{\sum_{j=1}^Kp_j^{\min}+\frac{\sigma^2}{h_m}-p_m^{\min}}
\nonumber\\
=\arg\min_{m\in\mathcal K}&
\frac{\sigma^2}{h_m}-p_m^{\min}.
\end{align}
Substituting \eqref{al2eq1} to \eqref{AppTh2eq2}, we have
\begin{align}
k&=\arg\min_{j\in\mathcal K}\frac{\sigma^2}{h_j}-\left(1-2^{\frac{a_j-R_j}{B}}
\right)
\left( P-p_0+\frac{\sigma^2}{h_j}
\right)
=\arg\min_{j\in\mathcal K} 2^{\frac{a_j-R_j}{B}}
\left( P-p_0+\frac{\sigma^2}{h_j}
\right).
\end{align}

This completes the proof of Theorem 2.

\section{Proof of Lemma 4}
\setcounter{equation}{0}
\renewcommand{\theequation}{\thesection.\arabic{equation}}

Denote the objective function (\ref{al2min3_2}) as $f(\boldsymbol a)$ and we have
\begin{align}
\frac{\partial^2 f(\boldsymbol a)}{\partial a_m a_k}
=
0, \quad \forall m \in\mathcal K,
\end{align}
\begin{align}
\frac{\partial^2 f(\boldsymbol a)}{\partial a_m^2}
=
\frac{(\ln2)u_m\left(d-\sum_{j=1,j\neq m}^K  u_j \right)}
{B\left(d-\sum_{j=1}^K  u_j \right)^2}, \quad \forall m\in\mathcal K,
\end{align}
and
\begin{align}
\frac{\partial^2 f(\boldsymbol a)}{\partial a_m a_n}
=
\frac{(\ln2)u_m u_n}
{B\left(\sum_{j=1}^K  u_j+\frac{\sigma^2}{h_k}\right)^2},\quad \forall m,n\in\mathcal K, m\neq n,
\end{align}
where
\begin{equation}\label{AppLe4eq1}
u_j=  2^{\frac{a_j-R}{B}}
\left( P-p_0+\frac{\sigma^2}{h_j}
\right), u_k=0, \quad\forall j\neq k,
\end{equation}
and
\begin{equation}\label{AppLe4eq2}
d={\sum_{j=1,j\neq k}^K
\left( P-p_0+\frac{\sigma^2}{h_j}
\right)+\frac{\sigma^2}{h_k}}.
\end{equation}

Then, we can obtain Hessian matrix of (\ref{al2min3_2}) as
\begin{equation}
 \nabla^2  f(\boldsymbol a) =\frac{(\ln2)\left(\left(d-\sum_{j=1}^K  u_j \right)\text{diag}(\boldsymbol u)+\boldsymbol u^T\boldsymbol u\right)}
 {B\left(\sum_{j=1}^K  u_j+\frac{\sigma^2}{h_k}\right)^2}
\end{equation}
where $\boldsymbol u=[u_1,u_2,\cdots,u_K]$.
Since $u_j>0$ for all $j\neq k$ and $u_k=0$ according to \eqref{AppLe4eq1}, both $\text{diag}(\boldsymbol u)$ and $\boldsymbol u^T\boldsymbol u$ are positive semi-definite.
Consequently, the  Hessian matrix of (\ref{al2min3_2})  is positive semi-definite. In consequence, (\ref{al2min3_2}) is a convex function.
This completes the proof of Lemma 4.
\section{Proof of Theorem 4}
\setcounter{equation}{0}
\renewcommand{\theequation}{\thesection.\arabic{equation}}

We first show that $a_m^*\geq a_n^*$ for all $m<n$ for the optimal solution $\boldsymbol a^*$ of problem (\ref{al2min3_2}) with equal rate demand.
This can be proved by the contradiction method.
If there exists $m<n$ such that $a_m^*<a_n^*$, we can construct a new solution $\boldsymbol a'$ with $a_m'=a_n^*$, $a_n'=a_m^*$, $a_j'=a_j^*$ for $j\neq m, n$.
Then, we can obtain
\begin{align}\label{AppTh4eq1}
&\sum_{j=1}^K
\left(1-2^{\frac{a_j'-R }{B}}
\right)
\left( P-p_0+\frac{\sigma^2}{h_j}
\right)
\nonumber\\
=&\sum_{j=1}^K
\left(1-2^{\frac{a_j^*-R }{B}}
\right)
\left( P-p_0+\frac{\sigma^2}{h_j}
\right)
+\left(2^{\frac{a_m^*-R }{B}}-2^{\frac{a_n^*-R }{B}}\right)\left(  \frac{\sigma^2}{h_m}
-\frac{\sigma^2}{h_n}
\right)
\nonumber\\
<&\sum_{j=1}^K
\left(1-2^{\frac{a_j^*-R }{B}}
\right)
\left( P-p_0+\frac{\sigma^2}{h_j}
\right),
\end{align}
where the inequality follows from the fact that $h_m\leq h_n$ and $a_m^*<a_n^*$.
Based on \eqref{AppTh4eq1}, we can claim that the new solution $\boldsymbol a'$ is feasible with better objective value then solution $\boldsymbol a^*$, which contradicts the fact that $\boldsymbol a^*$ is the optimal solution.

Then, we show that the objective function (\ref{al2min3_2}) monotonically increases with $a_j$ for all $j\neq k$.
To show this, the first derivative of the objective function (\ref{al2min3_2}) with respect to $a_j$ can be presented as:
\begin{align}\label{AppTh4eq2}
\frac{\partial f(\boldsymbol a)}{\partial a_j}
&=-1+\frac{ 2^{\frac{a_j-R}{B}}
\left( P-p_0+\frac{\sigma^2}{h_j}
\right) }
{\sum_{j=1,j\neq k}^K \left(1-2^{\frac{a_j-R}{B}}
\right)
\left( P-p_0+\frac{\sigma^2}{h_j}
\right)+\frac{\sigma^2}{h_k}}
\nonumber\\
&\geq -1+\frac{ 2^{\frac{a_k-R}{B}}
\left( P-p_0+\frac{\sigma^2}{h_k}
\right) }
{\sum_{j=1,j\neq k}^K \left(1-2^{\frac{a_j-R}{B}}
\right)
\left( P-p_0+\frac{\sigma^2}{h_j}
\right)+\frac{\sigma^2}{h_k}}
\nonumber\\
&=\frac{ P-p_0-\sum_{j=1}^K \left(1-2^{\frac{a_j-R}{B}}
\right)
\left( P-p_0+\frac{\sigma^2}{h_j}
\right)-\frac{\sigma^2}{h_k}
}
{\sum_{j=1,j\neq k}^K \left(1-2^{\frac{a_j-R}{B}}
\right)
\left( P-p_0+\frac{\sigma^2}{h_j}
\right)+\frac{\sigma^2}{h_k}}
\nonumber \\
&\geq 0,
\end{align}
where $f(\boldsymbol a)$  denotes the objective function (\ref{al2min3_2}).
The first inequality in \eqref{AppTh4eq2} follows from constraint
(\ref{al2min3_2}b), and the second inequality in \eqref{AppTh4eq2} follows from constraint
(\ref{al2min3_2}c).

Based on  \eqref{AppTh4eq2}, 
the objective function (\ref{al2min3_2}) increases with each rate $a_j$,
while \eqref{AppTh4eq1} shows that it is optimal to allocate rate to the user that has the lowest channel gain.
As a result, the optimal rate allocation can be given in
\eqref{al2th4eq1} and \eqref{al2th4eq2}.
This completes the proof of Theorem 4.


\section{Proof of Lemma 6}
\setcounter{equation}{0}
\renewcommand{\theequation}{\thesection.\arabic{equation}}

According to Theorem 4, we consider the following three situations of the optimal rate allocation.

1) If $\lfloor \frac{c_1}{R} \rfloor<1$,
we can obtain that the optimal rate allocation is $a_1^*=c_1$ and $a_2^*=0$.
Substituting $a_1^*=c_1$ in \eqref{sys1eq5} and $a_2^*=0$ to problem \eqref{al2min3} with $K=2$ and $R_1=R_2=R$ yields
\begin{subequations}\label{AppLe6eq1}
\begin{align}
\mathop{\max}_{ p_0 }\quad& R
+B\log_2  \left(\!1\!+\!\frac{
\left( P+\frac{\sigma^2}{h_1}
\right)2^{\frac{-R}{B}}-\frac{\sigma^2}{h_1}}
{  P-p_0+\frac{\sigma^2}{h_1}
- \left( P+\frac{\sigma^2}{h_1}
\right)2^{\frac{-R}{B}}
+\frac{\sigma^2}{h_2}}\right),\tag{\theequation}
\\
\textrm{s.t.}\quad \:
&P-p_0+\frac{\sigma^2}{h_1}
- \left( P+\frac{\sigma^2}{h_1}
\right)2^{\frac{-R}{B}}
+\left(1-2^{\frac{-R}{B}}
\right)
\left( P-p_0+\frac{\sigma^2}{h_2}
\right) \leq P-p_0,\\
&B\log_2 \left(\frac{h_1P  +\sigma^2}{h_1(P-p_0)+\sigma^2}\right)  \leq R,\\
&p_0\geq \frac P 2 + \frac{\theta +\sigma^2}{2h_1}.
\end{align}
\end{subequations}
According to (\ref{AppLe6eq1}a), we have
\begin{align}
&\frac{\sigma^2}{h_1}
- \left( P+\frac{\sigma^2}{h_1}
\right)2^{\frac{-R}{B}}
\leq -\left(1-2^{\frac{-R}{B}}
\right)
\left( P-p_0+\frac{\sigma^2}{h_2}
\right)\leq0,
\end{align}
which indicates that the objective function (\ref{AppLe6eq1}) monotonically increases with $p_0$.
As a result, the optimal $p_0$ of problem (\ref{AppLe6eq1}) is the maximum feasible $p_0^*$, i.e., constraint (\ref{AppLe6eq1}b) holds with equality, which can be given by
\begin{equation}
p_0^* =\left(1-2^{\frac{-R}{B}}
\right)
\left( P+\frac{\sigma^2}{h_1}
\right).
\end{equation}
In order to make sure that problem (\ref{AppLe6eq1}) is feasible, we must have
\begin{equation}
p_0^* \geq \max\left\{\frac{\sigma^2}{h_2} - \frac{\sigma^2}{h_1} +\frac{  P}
 { 1-2^{\frac{-R}{B}}},\frac P 2  + \frac{\theta +\sigma^2}{2h_1}\right\},
\end{equation}
according to constraints (\ref{AppLe6eq1}a) and (\ref{AppLe6eq1}c).
%

2) If $1\leq \lfloor \frac{c_1}{R} \rfloor<2$, we can obtain that the optimal rate allocation is $a_1^*=R$ and $a_2^*=c_1-R$, which simplifies the problem in \eqref{al2min3} as
\begin{subequations}\label{AppLe6eq3}
\begin{align}
\mathop{\max}_{ p_0 } \:& B\log_2 \left(\frac{h_1P  +\sigma^2}{h_1(P-p_0)\!+\!\sigma^2}\right)
+B\log_2  \left( \frac{{h_2}(P-p_0)+ {\sigma^2} }
{  {\sigma^2}
 } \right),\tag{\theequation}
\\
\textrm{s.t.}  \:\:\:
&P-p_0+\frac{\sigma^2}{h_2}
-\frac{P-p_0+\frac{\sigma^2}{h_2}}
{P-p_0+\frac{\sigma^2}{h_1}}
\frac{ P+\frac{\sigma^2}{h_1}}
{2^{\frac{R}{B}}}
  \leq P-p_0,\\
&R\leq B\log_2 \left(\frac{h_1P  +\sigma^2}{h_1(P-p_0)+\sigma^2}\right)  < 2R,\\
&p_0\geq \frac P 2 + \frac{\theta +\sigma^2}{2h_1}.
\end{align}
\end{subequations}
Since $h_1\leq h_2$ and the first derivative of (\ref{AppLe6eq3}) with respect to $p_0$ is
\begin{equation}
\frac{1}{P-p_0+\frac{\sigma^2}{h_1}}-\frac{1}{P-p_0+\frac{\sigma^2}{h_2}}\leq 0,
\end{equation}
the objective function (\ref{AppLe6eq3}) decreases with power $p_0$.
The optimal $p_0^*$ can be obtained from constraints (\ref{AppLe6eq3}b) and (\ref{AppLe6eq3}c), which can be given by:
\begin{equation}\label{AppLe6eq6}
p_0^* =\max\left\{\left(1-2^{\frac{-R}{B}}
\right)
\left( P+\frac{\sigma^2}{h_1}
\right),\frac P 2 + \frac{\theta +\sigma^2}{2h_1} \right\}.
\end{equation}

3) If $  \lfloor \frac{c_1}{R} \rfloor \geq 2$, we can obtain that the optimal rate allocation is $a_1^*=R$ and $a_2^*=R$.
In this case,
problem \eqref{al2min3} becomes
\begin{subequations}\label{AppLe6eq5}
\begin{align}
\mathop{\max}_{ p_0 } \:& B\log_2 \left(\frac{h_1P  +\sigma^2}{h_1(P-p_0)+\sigma^2}\right)
+B\log_2  \left( \frac{{h_2}(P-p_0)+ {\sigma^2} }
{  {\sigma^2}
 } \right),\tag{\theequation}
\\
\textrm{s.t.}  \:\:\:
& B\log_2 \left(\frac{h_1P  +\sigma^2}{h_1(P-p_0)+\sigma^2}\right)  \geq 2R,\\
&p_0\geq \frac P 2 + \frac{\theta +\sigma^2}{2h_1},
\end{align}
\end{subequations}
where the objective value also decreases with $p_0$.
Compared \eqref{AppLe6eq5} and \eqref{AppLe6eq5}, we can claim that the optimal $p_0^*$ should be achieved as the minimum value, i.e., \eqref{AppLe6eq6}.

According to the above three cases, it is observed that $a_1^*=R$.
Since the objective values (\ref{AppLe6eq1}), (\ref{AppLe6eq3}) and (\ref{AppLe6eq5})   do not change with $a_2$, we can show that $a_2^*=0$ is also optimal.
This completes the proof of Lemma 6.
\bibliographystyle{IEEEtran}
\bibliography{IEEEabrv,MMM}

\end{document}